# The Role of Plasma Shielding in Double-Pulse

# Femtosecond Laser-Induced Breakdown Spectroscopy

John S. Penczak,[a] Rotem Kupfer,[b] Ilana Bar,[b] and Robert J. Gordon[a1]

[a]Department of Chemistry, University of Illinois at Chicago, Chicago, IL 60680

[b]Department of Physics, Ben-Gurion University of the Negev, Beer-Sheva 84105, Israel

## Abstract

It is well known that optical emission produced by femtosecond laser-induced breakdown on a surface may be enhanced by using a pair of laser pulses separated by a suitable delay. Here we elucidate the mechanism for this effect both experimentally and theoretically. Using a bilayer sample consisting of a thin film of Ag deposited on an Al substrate as the ablation target and measuring the breakdown spectrum as a function of fluence and pulse delay, it is shown experimentally that the enhanced signal is not caused by additional ablation initiated by the second pulse. Rather, particle-in-cell calculations show that the plasma produced by the first pulse shields the surface from the second pulse for delays up to 100 ps. These results indicate that the enhancement is the result of excitement of particles entrained in the plasma produced by the first pulse.

---

[1] rjgordon@uic.edu

## I. Introduction

When an intense ultrashort laser pulse is focused onto a surface, a string of complex events is initiated, culminating in surface melting, plasma formation, ablation, and photoluminescence. In addition to being of fundamental interest for understanding laser-matter interactions, these processes have many practical applications, such as surface modification, high harmonic generation, and laser-induced breakdown spectroscopy (LIBS). The last of these processes provides a powerful analytical tool for quantitative elemental analysis of virtually any material, requiring little or no sample preparation.[1] Although the field has been dominated by the use of nanosecond (ns) lasers, femtosecond (fs) laser excitation has the advantages of greater sensitivity and improved spatial resolution.[2,3] These advantages stem from the property that fs laser radiation is absorbed much faster than thermal and mechanical relaxation. The smaller heat-affected zone is a consequence of the direct vaporization of the target, as opposed to ablation occurring through multiple steps, relying on melting and other thermal processes initiated by ns pulses.[4] Direct vaporization also makes the ablation process more stoichiometric as compared with ns-LIBS, where the pulse duration is long enough to interact with the plasma, ablation plume, and the thermally-affected zone.[5]

Previous work has shown that the sensitivity of fs-LIBS may be further enhanced by splitting the laser pulse into two sub-pulses and delaying the arrival of the second pulse to maximize the signal/background ratio. This method has been applied to a wide variety of materials, enhancing the signal/background ratio by a factor of ~2-50 as compared to that measured using a single ultrashort pulse of the same total energy.[6-10] Other studies, utilizing mass spectrometry, have found that double pulses (DPs) significantly increase the number density and kinetic energy of ions in the ablation plume.[4,11-15]

The goal of the present paper is to understand the mechanism responsible for the enhanced sensitivity of DP fs-LIBS. Three mechanisms have been proposed to explain the increase in the spectral line emission and ion yield, which is sometimes accompanied by a decrease in ablation efficiency.[16,17] The first mechanism hypothesizes that the first pulse melts the sample surface and that the second pulse interacts more strongly with the liquid layer.[6, 7, 13, 14] The second mechanism relies on the fact that the electronic thermal conductivity is proportional to the ratio of the electron and lattice temperatures, such that, as the two components equilibrate, the depth of the thermal energy diffusion decreases.[4] The resulting confinement of the absorbed energy leads to an enhanced signal produced by the second pulse. In the third mechanism, the first pulse creates plasma containing the species of interest. This plasma absorbs the second pulse, which excites the entrained particles.[12,15,18,19]

The first two mechanisms postulate that the enhanced signal is generated in the target itself, whereas the third mechanism assumes that the enhanced signal results from reheating of material entrained in the plasma above the surface. Previous attempts to distinguish between surface and plasma-centered mechanisms relied on measurements of the depth of the ablation crater. Such experiments are inconclusive, however, because recondensation of material from the plume may lead to ambiguous results.[20] Instead, we employ here the strategy of ablating a layered composite material with a pair of pulses under conditions such that the first pulse has insufficient energy to penetrate the outer layer, whereas a single pulse (SP) of the same total energy as the pulse pair can penetrate that layer to reach the substrate. Under these conditions, if the second pulse reaches the surface it may pool its energy with the first one to ablate the underlying layer and produce a LIBS signal from the substrate, whereas if the primary effect of the second pulse is to reheat the plasma, no LIBS signal is expected from the substrate. In addition, we use particle-in-cell (PIC) calculations to model the expansion of the plasma generated by the first pulse to determine under what conditions it can shield the surface from the second pulse.

## II. Experimental Methods

The experimental setup for this study, depicted in Fig. 1, is similar to that used in previous studies.[6,18] Briefly, pulses generated by a Ti:Sapphire laser (800 nm, ~60 fs duration) are directed through a Mach-Zehnder interferometer to create a pulse pair with a maximum delay of 104 ps. The properties of each sub-pulse were independently controlled using half-wave plates to change the polarization and variable neutral density filters to adjust the pulse energy in each arm of the interferometer. The laser was focused by a 10x, microscope objective lens (numerical aperture $NA = 0.25$), producing a focal diameter of $3.6 \pm 0.4$ μm. The pulse energy in these experiments ranged from 1 to 100 μJ, corresponding to fluences of ~20-2,000 J/cm$^2$ at the laser focus. The sample was mounted on a sub-μm precision xyz-translation stage, which was computer-controlled to move 100 μm between each shot. Each data point is an average of 10 shots. The laser was incident on the surface under ambient atmospheric conditions at a 30$^o$ angle with respect to the normal, and the detector was aligned perpendicular to the laser beam. Light emitted from the laser-induced plasma was collected and focused onto the 150 μm slit of a spectrograph (Spectrapro 2300i, Princeton Instruments), equipped with a thermoelectrically cooled charge coupled device (CCD) detector. The scattered laser light was blocked using a notch filter (Chroma Technology, E690SPUV6).

The ablation target consisted of a layer of Ag deposited on an Al surface by physical vapor deposition. The sample, fabricated at the Nanotechnology Core Facility at the University of Illinois at Chicago, had an outer layer uniformity of several nm. The thickness of this coating was chosen to satisfy the conditions described in the Introduction.  It was found empirically that a 500 nm layer of Ag enabled us to observe only Ag spectral lines at low fluence, while at higher fluences Al lines could also be detected.

## III. Experimental Results

Figure 2 shows data from a SP experiment using the 500 nm Ag/Al bilayer.  In this figure the integrated intensities of the strongest atomic emission lines (see below) are plotted as a function of fluence.  It is evident that the LIBS signal for Al is very weak for fluences below ~1,000 J/cm$^2$, whereas the Ag signal varies linearly with fluence over the entire measured range for both s- and p-polarized pulses.  In a separate control experiment the signal intensities of the same lines used in Fig. 2 were recorded using pure Al and Ag samples.  It was found for these bulk samples that the Al line is ~4 times more intense than the Ag line for SPs of equal fluence. Taking this calibration into account, the ratio of the Ag/Al line intensities produced from the bilayer sample can be calculated. At fluences capable of producing significant Al emission, the Ag/Al ratio decreases with fluence from ~300 at 490 J/cm$^2$ to ~50 at 1,965 J/cm$^2$.  At lower fluences the difference in signal intensity using either s- or p-polarized pulses is negligible.   At higher fluences, p-polarized pulses give a greater signal.  The difference in intensities for the two polarizations increases with fluence, with a maximum augmentation in signal strength for p-polarized pulses of 20% and 60% for Ag and Al, respectively, as compared to s-polarized pulses.

Examples of LIB spectra using SP or DPs with a total fluence of 1,965 J/cm$^2$ are shown in Fig. 3.  Here the delay of the DPs is set to its maximum value of 104 ps.  The electronic transitions involved are assigned from the NIST atomic spectra database[21] and are listed in the first four rows of Table I.  It is seen that DPs increase the total signal of the LIB spectra, including the background Bremsstrahlung continuum, yielding an average increase of the signal-to-background ratio by a factor of 2 for DPs compared to SPs of the same total energy. This result is in good agreement with experiments performed by previous groups at high fluences.[6,7,10]

Figure 4 shows a more dramatic enhancement effect, characteristic of the lower fluence regime, for SP vs. DPs with s- and p-polarization. The additional Ag lines, barely visible in Fig. 3, are identified in the last three rows of Table I.[21] The enhancement ratio (ER), defined as the background-corrected signal created by DPs divided by that of SPs, ranges from 2 to 4, depending on the spectral line and laser polarization. We find that p-polarized laser excitation gives a higher ER than s-polarization. This finding is in contrast with the SP case at low fluence, where s- and p-polarized excitations give nearly identical signal intensities at all wavelengths. The advantage of using DPs is clearly illustrated in the quality of the LIB spectra, where the atomic emission peaks are much more easily distinguishable against the continuum background signal.

In an attempt to elucidate the mechanism for DP enhancement, the ER was measured for the Ag/Al bilayer as a function of delay between the pulses. Figure 5 shows the result of this measurement using both p- and s-polarized pulses at a fluence of 588 J/cm$^2$. The data show a clear trend of a rapidly increasing enhancement up to 20-30 ps, at which time the ER reaches a plateau, changing very little from 40 ps to the maximum measured delay of 104 ps. This trend is observed for both Ag and Al and is in good agreement with previous studies by Gordon and co-workers[6] and by Pinon *et al.*[8] Again, p-polarized pulses show a distinctively higher ER than s-polarized ones. For delays less than 20 ps, the difference in ER using p- vs. s-polarized pulses is a monotonically increasing function of delay. In the long delay limit, p-polarized pulses show a nearly 50% increase in ER for both Ag and Al as compared to s-polarized pulses. It is also seen that Al has a slightly higher ER than Ag, with a difference ranging from 15-25% at pulse separations greater than 20 ps. Fitting of the ER to the function $A(1 - e^{(-|t|/\tau)})$ , as in ref. 6, where $t$ is the delay between the two laser pulses, gives a time constant, $\tau$, of roughly 20 ps for both Al and Ag with p- and s-polarized pulses.

The dependence of the ER on fluence was also examined. Figure 6 shows the ER measured at the maximum delay of 104 ps using s-polarized pulses. (Similar results were obtained for p-polarized pulses.) For fluences less than 500 J/cm$^2$ it was not possible to measure accurately the ER for Al because its signal intensity was barely above the background level. In the lower fluence regime the ER for Ag rapidly drops from a maximum value of ~9 at the lowest fluence of 49 J/cm$^2$ to ~2 at 500 J/cm$^2$, where it remains fairly constant at a value slightly less than 2 up to the highest fluence. This behavior is in good agreement with previous results.[7,9,10] The ER of Al is always slightly higher than that of Ag, but the difference is generally smaller at higher fluences.

The behavior of bulk Al differs in several respects from that observed for the bilayer, as shown in Fig. 7. At fluences less than ~100 J/cm$^2$, the qualitative trend in ER is similar to that of the bilayer, although its magnitude is only slightly greater than unity. The ER decreases with increasing fluence, and at the highest fluence measured in this experiment (490 J/cm$^2$) it was found that a SP actually produced *higher* line intensities than DPs by a factor of 1.13±0.07. To extend these measurements to still lower fluences, we replaced the objective with a simple focusing lens (f = 7.5 cm), having a spot size nearly 10 times that of the microscope objective. This weaker focus allowed measurement of the ER at fluences ranging from 2 to 25 J/cm$^2$. In this fluence regime there was no observable Al signal from the bilayer sample. The experiment on the bilayer using this lens may be interpreted, therefore, as an interaction with a bulk Ag target. This lens was also used to measure the line enhancement for a bulk Al sample. As shown in Fig. 8, the ER obtained with the simple lens attained a maximum value of ~10 and ~6 for Ag and Al, respectively, at the lowest fluence.  For the bulk Al sample, the ER decreased rapidly to less than a factor of 2 at 10 $J/cm^2$, whereas the ER for Ag remained high up to ~15 $J/cm^2$, finally converging to nearly the same value as that of Al at 25 $J/cm^2$.

In another experiment, the time constant,$\tau$, was measured for the bilayer as a function of fluence.  The result using the microscope objective is shown in Fig. 9.  As before, the low signal level for Al in the bilayer precluded measurement of $\tau$ at fluences below 500 $J/cm^2$. The time constant for Ag is ~30-40 ps for fluences below 250 $J/cm^2$ and falls to ~15 ps at 500 $J/cm^2$.  At higher fluences the time constants for both Al and Ag increase slowly to ~20 ps at 1,750 $J/cm^2$.

In a final experiment we compared the change in the LIBS signal for DPs separated by times greater than the entire ablation and equilibration cycle with that produced by a DP with a 104 ps delay.  For the first pulse pair, SPs with a fluence of 100 $J/cm^2$ (intensity $\approx 1.7 \times 10^{15}$ $w/cm^2$) were launched separated by delays of several seconds.  The spectrum produced by each laser shot is shown in the bottom two traces of Fig. 10(a). At this low fluence, a SP is incapable of penetrating the Ag film and ablating the Al substrate, whereas the second pulse focused on the same spot shows an enormous increase in the Al signal at 396.15 nm, along with a decrease in the Ag lines. ERs of this magnitude were not seen for this Al line at any fluence.   For comparison, the spectra of SPs and DPs with a delay of 104 ps at a total fluence of 200 $J/cm^2$ are shown in the top two traces.  The SP, having a fluence of 200 $J/cm^2$, is above the threshold for producing an Al signal, as shown in Fig. 10(b).  For the DP, however, each component of the pulse pair lies below the Al threshold and no Al signal is observed.  Fig. 10(b) shows that the threshold for producing an Al signal from the bilayer is ~300 $J/cm^2$, and double pulse enhancement does not begin until ~325 $J/cm^2$.

## IV. Particle-in-Cell Simulations

The evolution of the plasma generated by the first pulse was simulated with a 2D finite-dimensions-time-domain[22] PIC[23,24] (FDTD-PIC) code[25] using the Yee algorithm.[26] In this code, Maxwell's equations were time-integrated[26] self-consistently, with the motion of macro-particles pushed according to the relativistic Lorentz force[27] and the corresponding current density.[28] All nonlinearities arising from the motion of the particles were inherently included in the calculation.[29]

The code was used to calculate the plasma density profile as a function of time. A 50 fs, 800 nm laser pulse with an intensity of $5 \times 10^{15}$ W/cm$^2$ (corresponding to a fluence of 250 J/cm$^2$) was launched upon a uniform, 400 nm wide plasma slab (slab density $n_{e0} = 6 \times 10^{22}$ cm$^{-3}$) at a 30$^\circ$ angle from the normal. The electromagnetic mesh constant was initially 20 nm, containing 50 macro-electrons and 50 macro-ions per cell to ensure a sufficient number of particles in the Debye sphere.[24,24] The particles were allowed to propagate continuously inside the mesh with sub-pm resolution. With these parameters the simulation was numerically stable, and further refinement of the electromagnetic mesh or an increase of the number of particles did not affect the results.

Figure 11 shows the calculated electron number density profile at different times after the laser pulse. As time progresses, the very steep initial density gradient relaxes so that $n_e$ becomes fairly uniform over a range of several μm. Of particular interest is the electron density just above the surface, $n_e(x = 0, t)$. As shown in Fig. 12, this value falls monotonically with time, and for t > 110 ps it falls below the critical density, $n_c$,[30]

$$n_c = \frac{m_e \omega_0^2}{4\pi e^2} = 1.1 \times 10^{21} / \lambda^2 \ cm^{-3}, \qquad (1)$$

where $m_e$ is the electronic mass, $e$ is the electronic charge, $\omega_0$ is the laser frequency, and λ is the laser wavelength in microns. Additional calculations showed that $n_e(x = 0, t)$ scales approximately linearly with the shelf density, $n_{e0}$. Plotting the coordinate of the critical density, x(n$_c$), as a function of time in Fig. 13, we see that the critical surface initially moves outward to a maximum distance of 3.5 μm and then collapses back to the surface at t ≈ 100 ps. Taking the derivative of this curve, we find that the velocity of the critical surface starts out at $6 \times 10^4$ m/s at t = 0 and falls to zero at t ≈ 60 ps. Another set of calculations performed for an initial plasma slab having a density of $1 \times 10^{23}$ cm$^{-3}$ and a thickness of 100 nm and a laser intensity of $1 \times 10^{15}$ W/cm$^2$, which is closer to the conditions of Fig. 10(a), gave similar results except that the maximum distance of the critical surface extended 1 μm further and the plasma density dropped below the critical value 10 ps earlier.

## 4. Discussion

The experiments and calculations presented here provide definitive evidence that the DP enhancement of the LIBS signal of Ag at fluences ≥ 100 J/cm$^2$ is caused by plasma heating. One piece of experimental evidence is displayed in Fig. 10. In the bottom trace of this figure, a SP penetrates more than 50% of the Ag layer without reaching the substrate. If a second pulse of the same energy is launched after ablation by the first pulse has been completed and the surface has reached equilibrium, that pulse penetrates the remaining Ag layer and produces a strong Al signal. If, however, the second pulse arrives after a delay of only 104 ps, no Al signal is detectable. The absence of an Al signal in this case indicates that the second pulse was unable to penetrate the Ag layer and was absorbed by the plasma generated by the first pulse.

A second piece of evidence is provided by the time constant for the ER plotted in Fig. 9. This time constant is associated with a transport process to be discussed later in the paper. The point we wish to make here is that if the enhancement occurs homogeneously (i.e., the Ag and Al atoms or ions excited by the second pulse are in the same volume when the second pulse arrives), we would expect τ to be the same for both, whereas if the excited Al species is located in an underlying layer we would expect it to have a longer time constant than Ag. The observation of identical values of τ for the two metals over a wide range of fluences rules out enhancement at the surface of the bilayer structure.

A third piece of evidence comes from the PIC calculation of the plasma expansion. The frequency-dependent index of refraction of the plasma is given by[31]

$$n(\omega) = \left(1 - \frac{n_e}{n_c}\right)^{1/2},$$ (2)

where $n_c$ is given by Eq. (1). It follows that an over-dense plasma ($n_e > n_c$) reflects and absorbs but does not transmit light of frequency $\omega$. Further analysis[31] shows that p-polarized radiation incident at an angle $\theta$ to the normal is reflected at a location in the plasma where $n_e = n_c \cos^2 \theta$. A component of the incident wave tunnels into the plasma evanescently until it reaches the critical surface (i.e., at $n_e = n_c$), where it may be absorbed resonantly. S-polarized radiation may also be absorbed by the plasma by a mechanism called collisional absorption or inverse Bremsstrahlung, with maximum absorption occurring at the critical density. The collision frequency decreases with plasma temperature, reducing the contribution of collisional absorption at high fluences.[30]

The principle conclusion to be drawn from the PIC calculations is that for delay times less than ~100 ps the electron number density is high enough to shield the surface from the second pulse. This finding implies that for times $\leq 100$ ps any observed enhancement of the LIB signal must be generated in the plasma itself. The behavior at t > 100 ps is a moot point because we cannot determine from the present study the opacity of neutral particles (including nano-structures) in the plume, which might shield the surface after the electron density drops below $n_c$.

We also cannot determine from the present study whether plasma shielding plays a role at much lower intensities. We expect for most materials that LIB generates a plasma on the surface that is initially over-dense.[32-34] Additional experiments with thinner bilayers and accompanying

PIC calculations could determine whether a crossover to the liquid surface mechanism might occur at lower intensity.

The plasma heating mechanism also explains the polarization dependence of the emission enhancement because resonant absorption of the second pulse is possible only for p-polarized light.[31] The comparable signals observed for both polarizations of SPs at fluences below 500 J/cm$^2$ suggest that collisional absorption dominates in that regime. We note, however, that the absorption of the s-polarized light may also be explained by rippling of the critical surface caused by a Rayleigh-Taylor-like instability.[35-37]

The falloff of the ER with fluence shown for Ag in Fig. 6 and values of ER < 1 for bulk Al shown in Fig. 7 are consistent with the plasma-shielding mechanism. Both sets of observations may be explained if the fraction of electronically excited atoms and ions produced by the first pulse increases with fluence. Since the second pulse cannot penetrate the plasma shield, as the fraction of unexcited particles decreases so must also the ER.

It is instructive at this point to examine the three models mentioned in the Introduction to see how they conform to the data reported here. The liquid absorption model assumes that the liquid layer generated by the first pulse absorbs energy more efficiently than the solid, producing a higher energy density and therefore higher temperatures within the material.[14] These higher temperatures would then cause an increase in both the amount of material removed and the concentration of excited species, giving an overall enhancement of the photoluminescence. The liquid absorption mechanism argues that the dependence of the signal enhancement on the delay between the pulses is determined by the position of the melt front. After the first pulse, the high temperatures created at the surface propagate into the material at approximately 10% of the sonic velocity, melting deeper layers. The increase in enhancement with delay is a result of the laser having more liquid to interact with.  At some point the thickness of the liquid layer exceeds the optical penetration depth of the laser, and the enhancement levels off.[13]  The time constant, $\tau$, is expected to decrease with fluence because the sonic velocity increases with temperature, reducing the time it takes the melt front to penetrate the material. This model can describe the behavior of a semiconductor such as Si, for which the optical absorption length of the liquid is $10^{-3}$ that of the solid.[6]  In the case of metals, the model may be modified by taking the appropriate scale length to be the penetration distance of ballistic electrons, which is greater for the solid.[18]  The present experiment shows that, at least for Ag, absorption of the second pulse by the plasma at high fluence rules out the liquid surface model.

For Si, the liquid absorption model was supported by the observation that the crater depth produced by a pair of pulses, each having energy E, was twice that produced by a SP of energy E.[6,20]  It was also observed, however, that the crater depth varied little with pulse delay.  This observation was reconciled with the model by noting that the crater contained amorphous material that increased with pulse delay.  These observations were interpreted to mean that the second pulse reached the surface and enhanced both the amount of ablation and the LIBS signal, and that at later times the crater was partially filled in by material condensed from the plume.

The results of the Si experiment may need to be reinterpreted in light of the present findings. As shown in Fig. 10, at low fluences a DP with a delay of 104 ps does not produce a LIB signal from the underlying Al layer, whereas if the delay is on the order of seconds, after the liquid has solidified, a large Al signal is produced.  The implication is that for short delays the second pulse does not reach the surface.  The greater ablation depth for a DP may be caused by a shock wave launched by the second pulse at the critical surface of the plasma.  Material carried by this shock wave may be responsible for the amorphous material deposited in the crater.

Likewise, a scanning electron microscope image of a crater produced in Si by a DP at low fluence,[13] which was cited as evidence for the second pulse coupling directly with a molten surface, may also be interpreted as redeposition of amorphous material in the crater.

The time constant for the ER for Si may also need to be re-interpreted.[18] For the liquid absorption model, $\tau$ is the time it takes the melt front or the ballistic electrons to propagate to the optical skin depth. This interpretation is inconsistent with the present results, where $\tau$ is the same for Ag and Al. If the liquid absorption model were correct, the Ag signal would cease growing once the melt front reached the interface, whereas the Al emission would continue to grow after the melt front has passed the Ag/Al interface.

The second mechanism, which has received much less attention, is based on the dependence of the electronic thermal conductivity, $k_e$, on the ratio of the temperatures of the electrons, $T_e$, and the lattice, $T_l$. Hohlfeld $et$ $al.$[38] showed that $k_e$ decreases as the electron and lattice temperatures equilibrated. Noel $et$ $al.$[4] postulated that at short delays, before the electron and lattice could equilibrate, the high $k_e$ allows the energy of the second pulse to penetrate more deeply. This effect would favor increased nanoparticle production and ablation efficiency. Conversely, as $T_e/T_l$ approaches unity, the electron thermal conductivity would be smaller and the laser energy would be more confined, resulting in a higher temperature gradient and favoring increased atomization and smaller ablation depths. The transition from nanoparticle production to atomization would result in higher optical emission and ion production.

The electron conductivity mechanism predicts that $k_e$ increases with fluence, reducing the energy density and optical emission. This prediction is in accord with our observation that the ER decreases with fluence. This model also predicts that the thermal equilibrium time increases with fluence, in contradiction with our observation in Fig. 9 that $\tau$ decreases with fluence in the lower intensity regime and increases only slightly in the higher intensity regime. As was the case for the liquid absorption model, our observation that the second laser pulse is absorbed or reflected by the plasma rules out the electron conductivity mechanism for the emission enhancement.

The final mechanism, known as the plasma reheating mechanism, was primarily invoked to explain the reduction in ablation efficiency of metals when using DPs separated by longer than a few ps. Multiple studies have shown that three distinct timescales govern the ablation efficiency of DPs.[12,15,16,18,19] For delays less than a few ps, the resulting ablation crater depth is equal to that created by a SP of same total energy. This finding is reasonable because the electron-lattice equilibration time, $\tau_{el}$, is on the order of a few ps in metals. This means that the energy of the first pulse had not been dissipated, and no significant expansion or ablation had occurred by the time the second pulse arrived, such that the energy of the second pulse was absorbed as if the pulses had arrived simultaneously. In the second time domain, it was found for Cu[16] and Ag[17] that the ablation depth for DPs decreased monotonically with increasing delay up to 10-20 ps, at which point the ablation depth for a DP of energy 2E was equal to that created by a SP of energy E. This indicated that only the first pulse was responsible for ablating the material and that the energy of the second pulse never reached the surface, either being reflected or absorbed by the plasma created by the first pulse. Some studies have shown that the ablation depth of DPs, separated by more than 10-20 ps, had a slightly smaller ablation depth than a SP of half the total energy.[15,39] This result was explained by Povarnitsyn *et al.*[39] as caused by interference between a compression wave produced by the second pulse with a rarefaction wave produced by the first pulse. It is also possible, however, that the reduction in the crater depth may have been caused by deposits of amorphous material, as was observed for Si.

If the plasma reheating mechanism is responsible for the emission enhancement, the time constant must be related to transport of neutral atoms into the critical density region. We conjecture that Ag and Al atoms ejected by the first pulse are carried by a shock wave.  Initially this shock wave trails the rapidly expanding electron cloud, but as the motion of the critical surface slows down and reverses direction (Fig. 13), the atomic and ionic species overtake it and may be excited by the second pulse.  (The reversal of the critical surface velocity is a consequence of the expansion and rarefaction of the plasma, so that the location of the critical surface eventually gets closer to the solid surface before it disappears entirely.)

## V. Conclusions

The experimental and theoretical results presented here show that enhancement of the fs-LIB signal produced by a pulse pair is caused by excitation of particles entrained in the plasma produced by the first pulse rather than by increased ablation of the surface by the second pulse. Using a thin layer of Ag deposited on an Al substrate as the ablation target, we demonstrated that if the first pulse does not penetrate down to the Al layer, the second pulse may be able to ablate the substrate only after a very long delay when the plasma and plume produced by the first pulse have completely dissipated. For delays less than 100 ps, the plasma produced by the first pulse shields the surface from the second pulse.  We also showed that the time constant associated with the enhanced LIB signal is the same for Ag and Al, indicating that the enhanced Al signal does not originate from material buried beneath the Ag coating.  The greater enhancement for p-polarized radiation is indicative of resonant plasma absorption, which is consistent with the plasma reheating mechanism.  These conclusions apply to ablation of Ag with fluences $\geq 100$ J/cm$^2$ (intensities $\geq 1.7 \times 10^{15}$ W/cm$^2$).  Additional experiments and calculations are required to determine whether there is a change in mechanism at lower intensities.

**Acknowledgements**

This project was supported by the U.S. Air Force Surgeon General's Office (AF/SG) under Contract Number FA7014-07-C-0047 and administered by the Air Force District of Washington (AFDW).

**Table I.**  Electronic transitions of aluminum and silver

| Species | Wavelength (nm) | Configuration | J |
|---------|-----------------|---------------|---|
| Al I | 394.40 | $3s^2 3p$ - $3s^2 4s$ | $1/2 - 1/2$ |
| Al I | 396.15 | $3s^2 3p$ - $3s^2 4s$ | $3/2 - 1/2$ |
| Ag I | 520.91 | $4d^{10} 5p$ - $4d^{10} 5d$ | $1/2 - 3/2$ |
| Ag I | 546.54 | $4d^{10} 5p$ - $4d^{10} 5d$ | $3/2 - 5/2$ |
| Ag I | 405.85 | $4d^{10} 5p - 4d^{10} 6d$ | $1/2 - 3/2$ |
| Ag I | 421.62 | $4d^{10} 5p - 4d^{10} 6d$ | $3/2 - 5/2$ |
| Ag I | 431.32 | $4d^{10} 5s 5p - 4d^9 5s 6s$ | $5/2 - 7/2$ |

**Figure Captions**

**Figure 1.** Schematic drawing of the apparatus including half-wave plates ($\lambda/2$), polarizers (P1 and P2), beam splitters (Bs1 and Bs2), variable neutral density filters (Vf2), interferometer mirrors (M1 and M2), and lenses (L1, L2, and L3).

**Figure 2.** LIBS signal acquired using single pulses on a 500 nm thick coating of Ag on Al, using s-polarized (green diamonds and red circles ) or p-polarized (blue triangles and black squares) pulses. The data correspond to the strongest spectral lines of Al (396.15 nm, lower two sets of points) and Ag (546.54 nm, upper two sets of points) in our wavelength range at nm. The signal for Ag is divided by a factor of 10 to facilitate a visual comparison between the two species.

**Figure 3.** LIB spectra of a bilayer sample using a p-polarized laser with a fluence of 1,965 J/cm$^2$. The black and red curves correspond to double pulses separated by 104 ps or a single pulse, respectively, with the same total energy. The transitions of the labeled peaks are listed in Table I.

**Figure 4.** LIB spectra for a bilayer taken with a fluence of 588 J/cm$^2$ with p-polarized (panel a) and s-polarized (panel b) using single (red curve) and double (black curve) pulses.

**Figure 5**. Enhancement ratio for double vs. single pulses of the signal intensity for the Ag line at 546.54 nm and the Al line at 396.15 nm.  The data were recorded using a bilayer at a fluence of 588 J/cm$^2$.   The data correspond to the use of p/s-polarized pulses for Al (black squares/blue triangles) and Ag (red circles/green triangles). The error bars are typical standard deviations.

**Figure 6.**  Enhancement ratio obtained from the bilayer sample at a delay of 104 ps as a function of fluence with s-polarized light.  The data points for Ag (red squares) and Al (black circles) at 588 J/cm$^2$ are the same as the 104 ps data in Fig. 5.  The error bars represent the standard deviation of the measurement.

**Figure 7.** Enhancement ratio (ER) for the Al line at 396.15 nm using a bulk Al sample.  The data were recorded at a fluence of 98 J/cm$^2$ (black squares), 196 J/cm$^2$ (red circles), and 490 J/cm$^2$ (blue triangles).  The asymmetry in these plots may be due to a slight misalignment of the interferometer, which is not noticeable on the larger scales of Figs. 5 and 6.

**Figure 8.** Enahncement ratio obtained at a delay of 104 ps as a function of fluence using a simple focusing lens.  The wavelength analyzed for Ag (red squares), using the bilayer sample, and Al (black circles), using bulk Al, are the same as in Fig. 5.  The error bars represent the standard deviation of the measurement.

**Figure 9.** Time constant from the fitting of the function $A(1 - e^{(-|t|/\tau)})$ to the enhancement ratio dependence on delay, $t$, as a function of fluence.  The same signals for Ag (red squares) and Al (black squares) were used as in Fig. 5.  The error bars represent the standard deviation for multiple shots.

**Figure 10.** LIB spectra of the Ag/Al bilayer sample. (a) The bottom two traces represent the signal acquired by a single shot of 100 J/cm$^2$ fired in the same spot separated by a few seconds. The top two curves represent a single pulse (SP) and double pulses (DP) with a 104 ps delay with a total fluence of 200 J/cm$^2$. The spectra are vertically offset to ease viewing. (b) Fluence dependence of the Al LIB signal for single and double pulses. The straight lines are least square fits of the LIB signal above threshold, drawn to guide the eye.

**Figure 11.** Particle-in-cell calculation of the electron number density profile for an initial slab density of 6x10$^{22}$ cm$^{-3}$ and a laser intensity of 5x10$^{15}$ W/cm$^2$.

**Figure 12.** Particle-in-cell calculation of the electron density produced by the first pulse just above the target surface, assuming the parameters of Fig. 11.

**Figure 13.** Particle-in-cell calculation of the location and velocity of the critical surface for the parameters of Fig. 11.

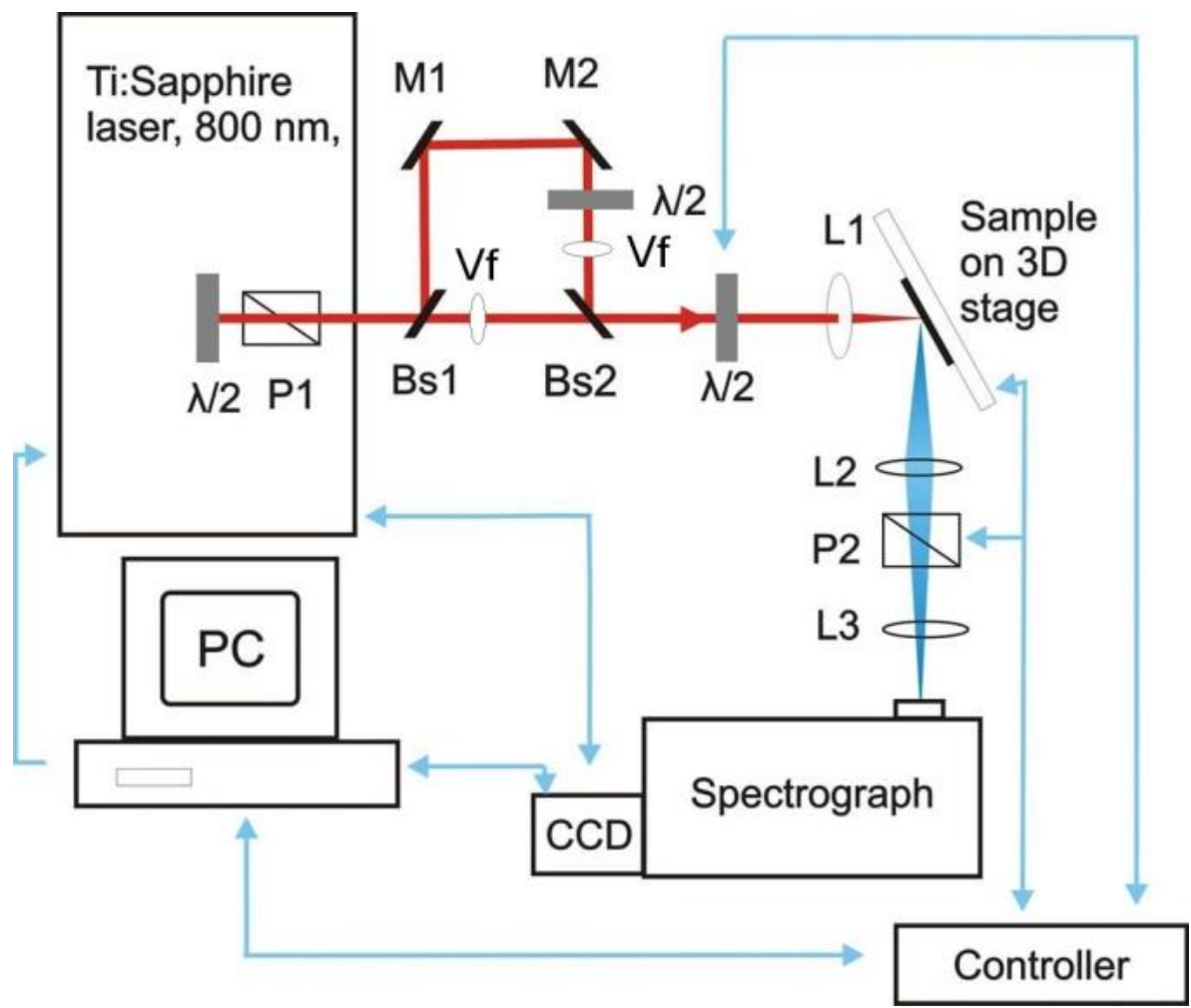

Figure 1

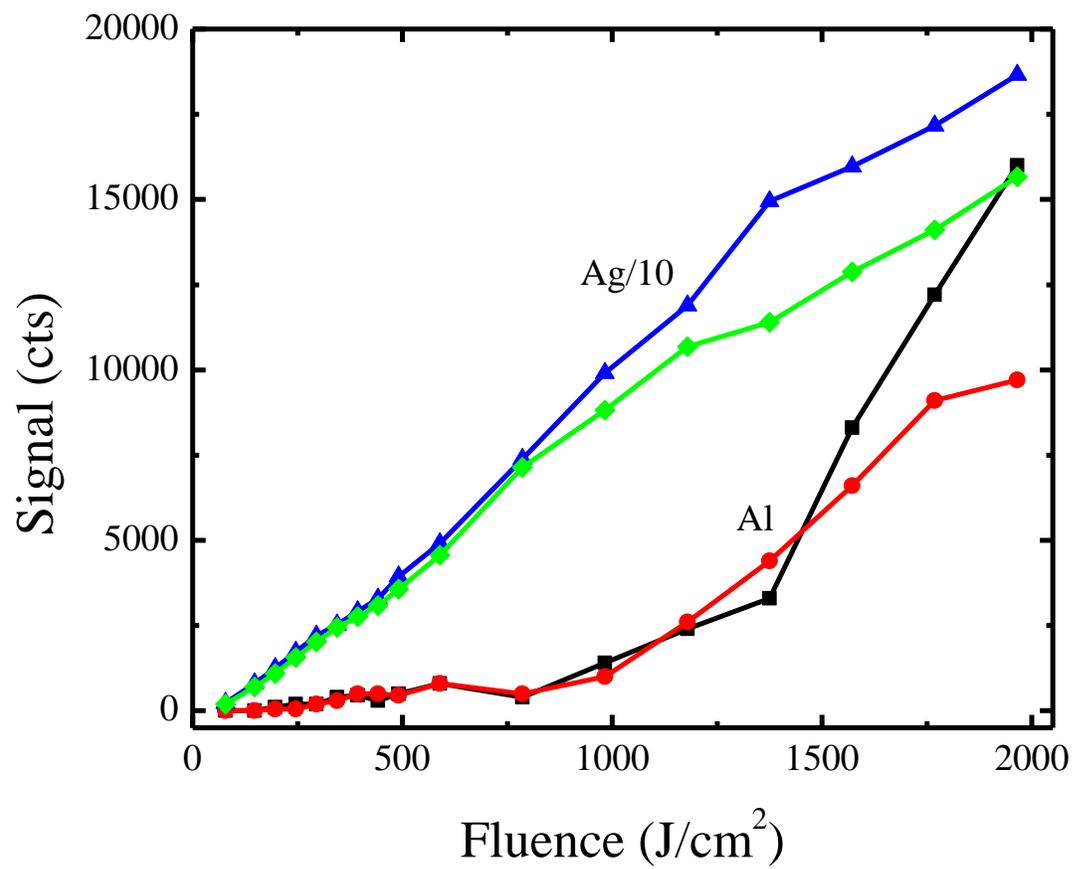

Figure 2

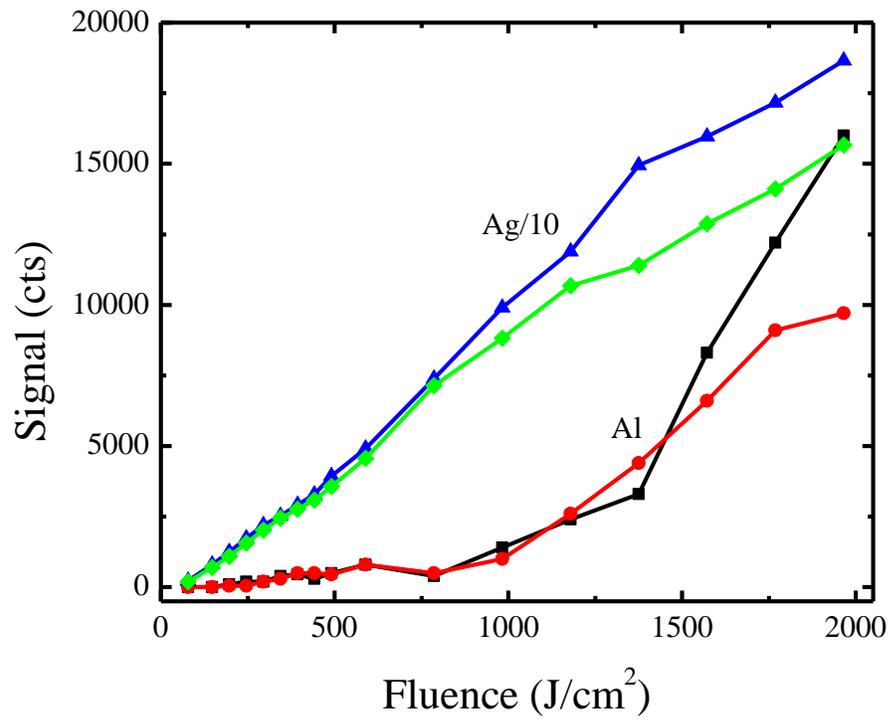

Figure 3

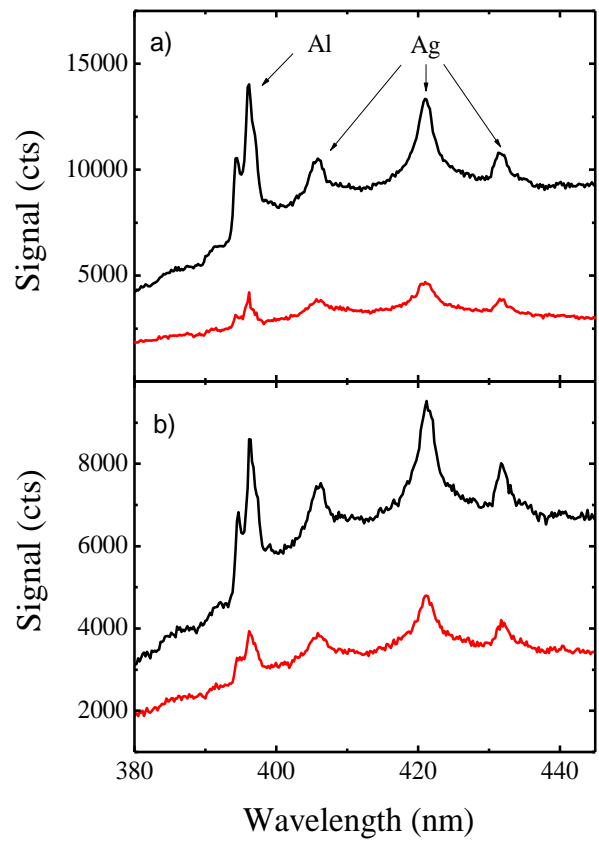

Figure 4

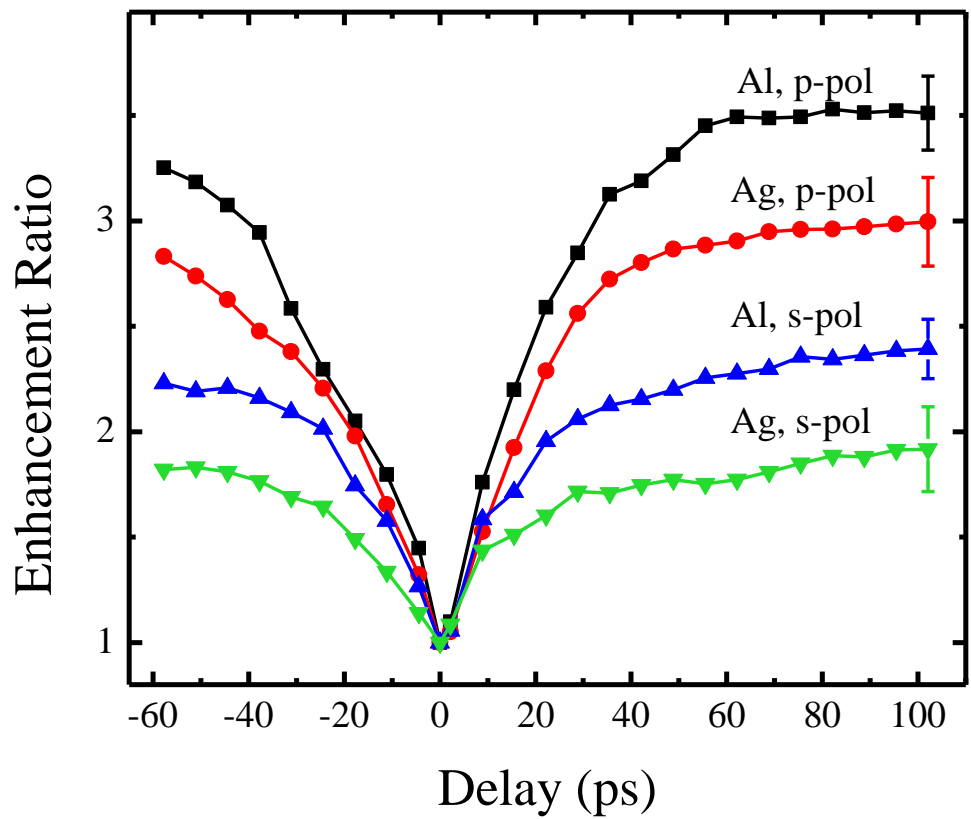

Figure 5

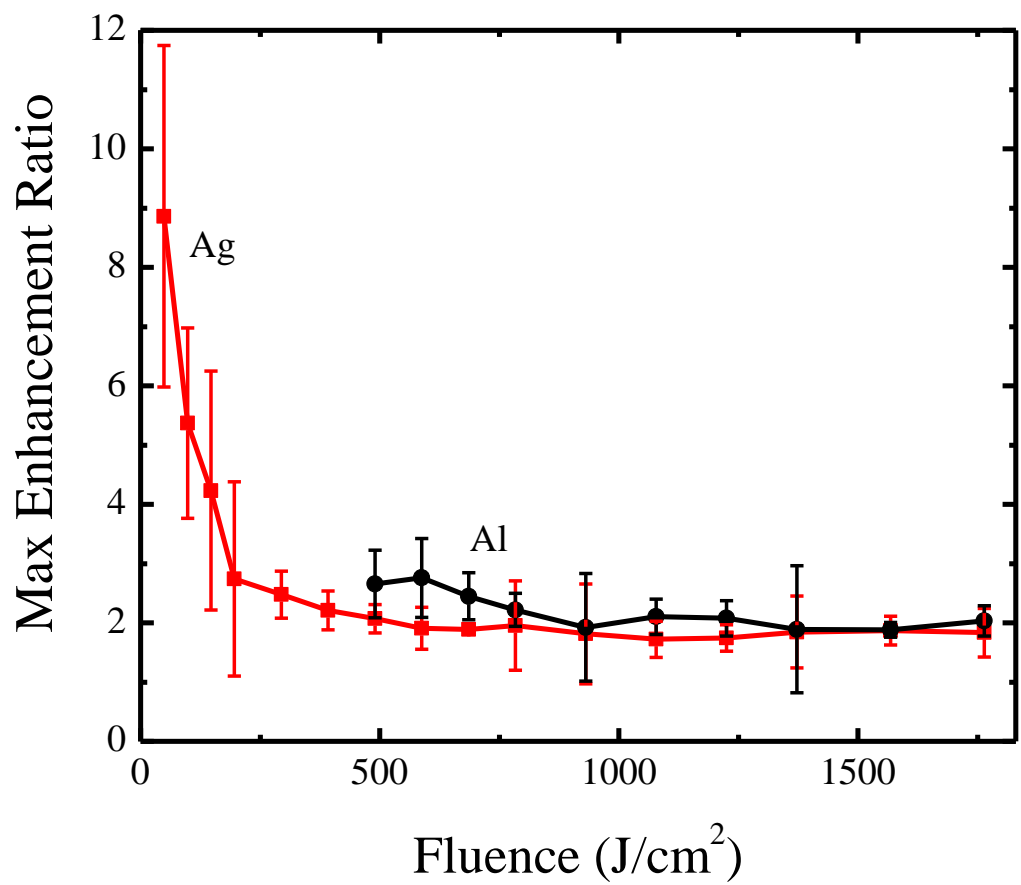

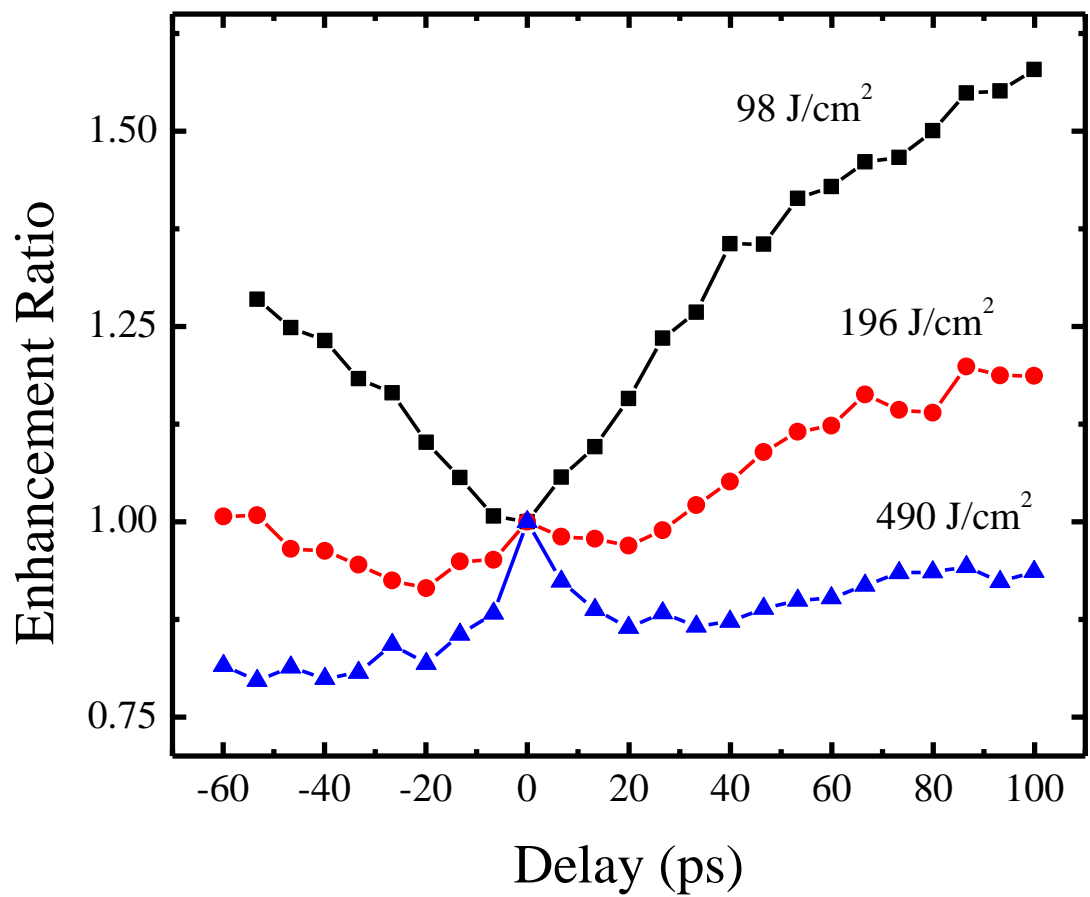

Figure 7

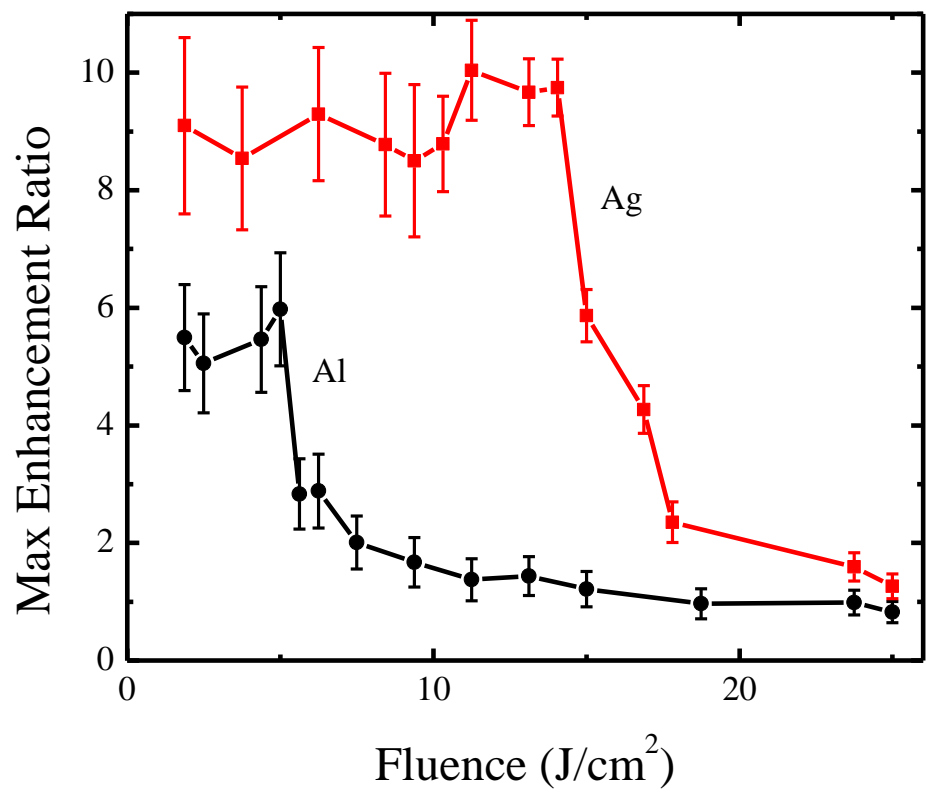

Figure 8

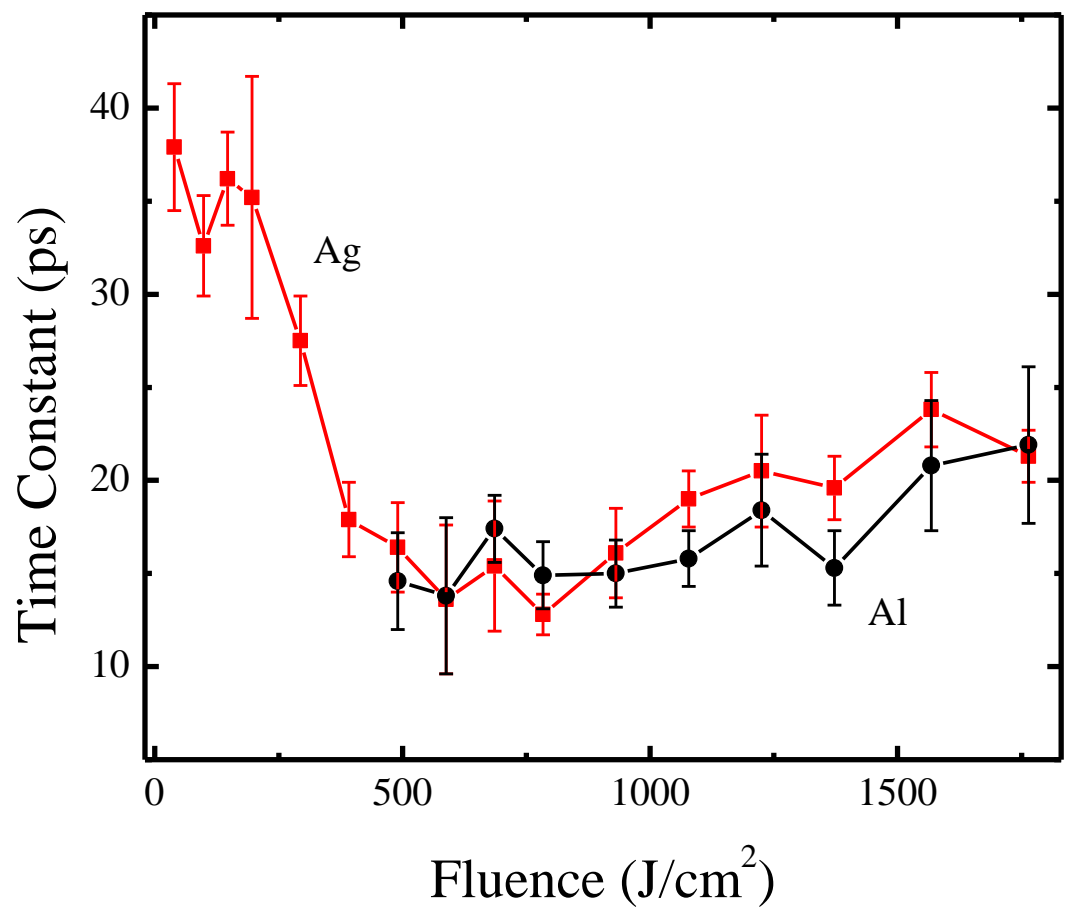

Figure 9

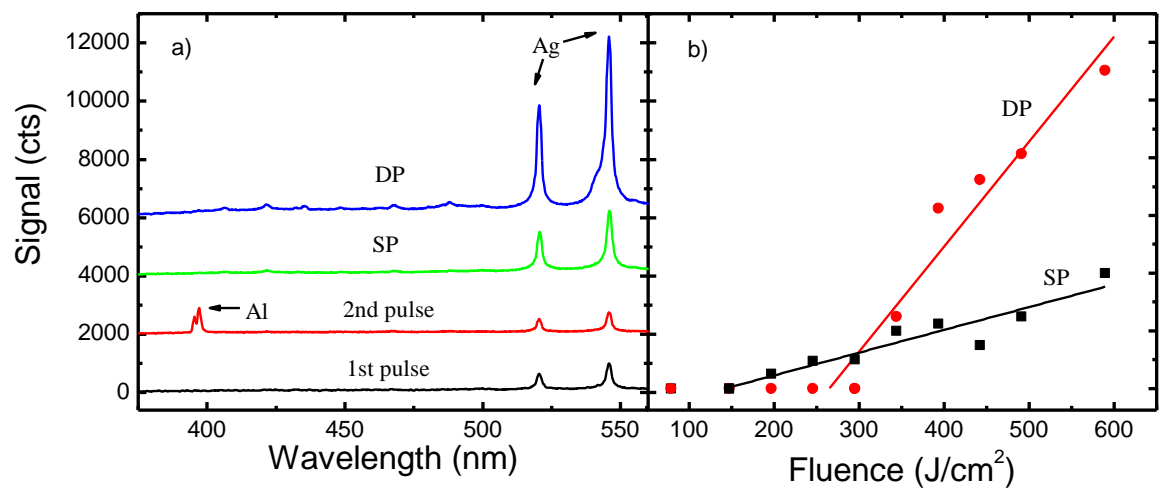

Figure 10

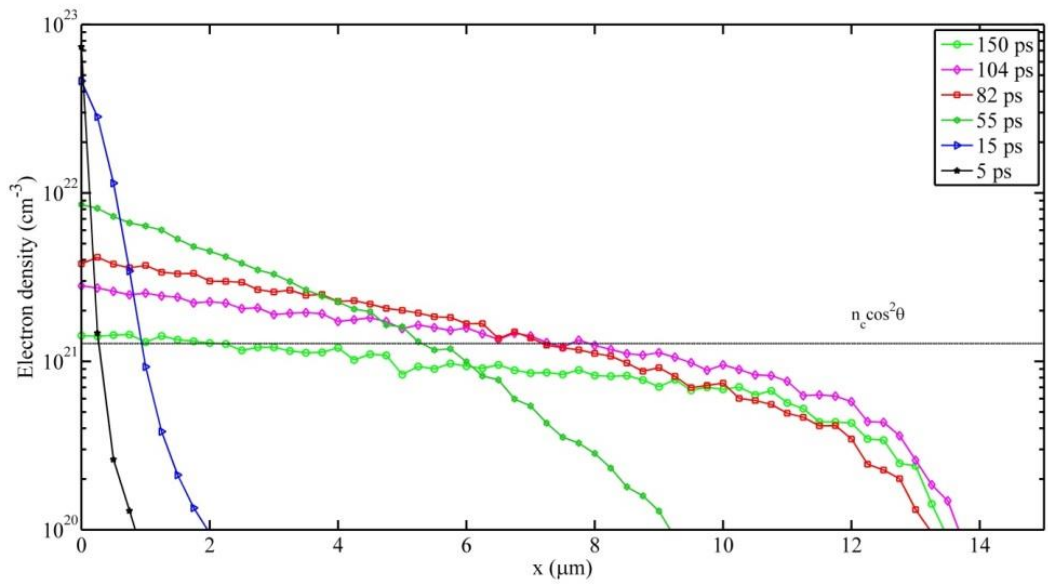

**Figure 11**

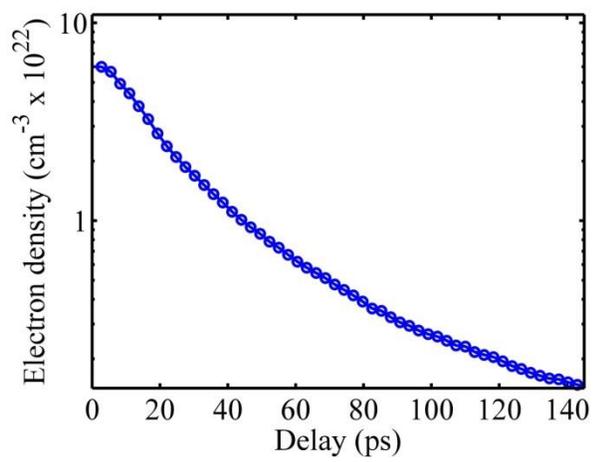

**Figure 12**

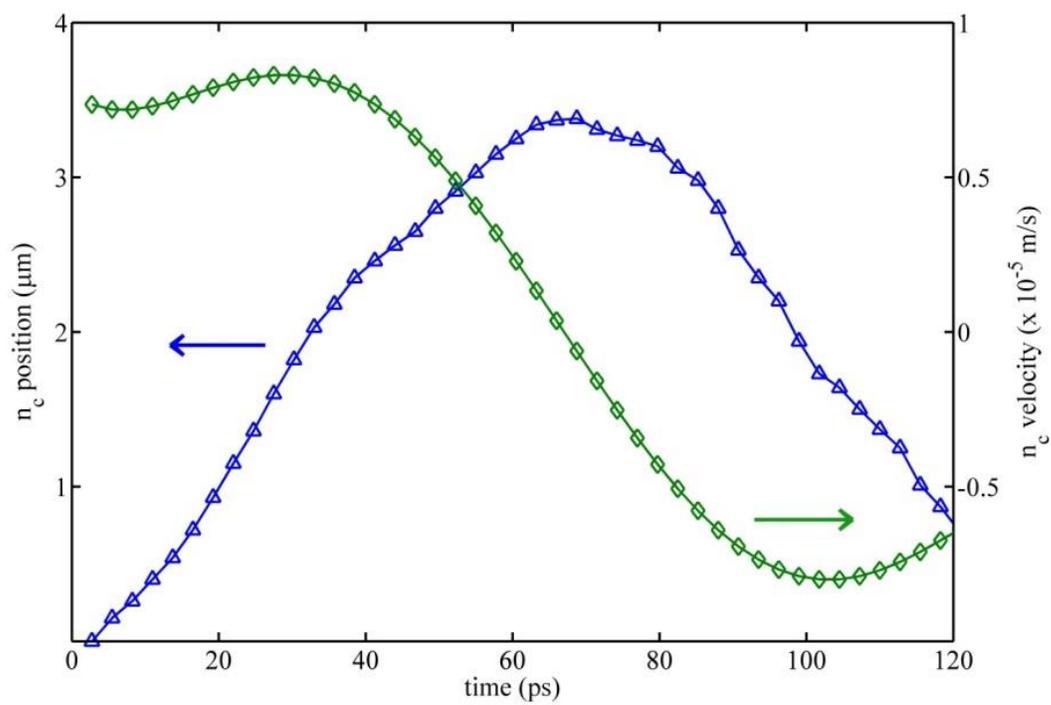

**Figure 13**

[1] D. A. Cremers and L. J. Radziemski, *Handbook of Laser-Induced Breakdown Spectroscopy*, (Wiley, New York, 2006).

[2] J. Scaffidi, J. Pender, W. Pearman, S. R. Goode, B. W. Colston Jr, J. C. Carter and S. M. Angel, Appl. Opt. **42**, 6099 (2003).

[3] K. L. Eland, D. N. Stratis, D. M. Gold, S. R. Goode and S. M. Angel, Appl. Spectrosc. **55**, 286 (2001).

[4] S. Noel, E. Axente and J. Hermann, Appl. Surf. Sci. **255**, 9738 (2009).

[5] R. E. Russo, X. L. Mao, C. Liu and J. Gonzalez, J. Anal. At. Spectrom. **19**, 1084, 2004.

[6] Z. Hu, S. Singha, Y. Liu and R. J. Gordon, Appl. Phys. Lett. **90**, 131910 (2007).

[7] S. Singha, R. J. Gordon and Z. Hu, Chin. Phys. Lett. **25**, 2653 (2008).

[8] V. Pinon, C. Fotakis, G. Nicolas and D. Anglos, Spectrochim. Acta Part B **63**, 1006 (2008).

[9] V. Pinon and D. Anglos, Spectrochim. Acta Part B **64**, 950 (2009).

[10] J. Guo, T. Wang, J. Shao, T. Sun, R. Wang, A. Chen, Z. Hu, M. Jin and D. Ding, Opt. Comm. **285**, 1895 (2012).

[11] V. I. Babushok, F. C. DeLucia Jr, J. L. Gottfried, C. A. Munson and A. W. Miziolek, Spectrochim. Acta Part B **61**, 999 (2006).

[12] S. Amoruso, R. Bruzzese, X. Wang, G. O'connell and J. G. Lunney, J. Appl. Phys. **108**, 113302 (2010).

[13] M. Spyridaki, E. Koudoumas, P. Tzanetakis, C. Fotakis and R. Stoian, Appl. Phys. Lett. **83**, 1474 (2003).

[14] E. Koudoumas, M. Spyridaki, R. Stoian, A. Rosenfeld, P. Tzanetakis, I. V. Hertel and C. Fotakis, Thin Solid Films **453-454**, 372 (2004).

[15] T. Donnely, J. G. Lunney, S. Amoruso, R. Bruzzese and X. Wang, J. Appl. Phys. **106**, 013304 (2009).

[16] A. Semerok and C. Dutouquet, Thin Solid Films **453-454**, 501 (2004).

[17]D. E. Roberts, A. du Plessis and L. R. Botha, Appl. Surf. Sci. **256**, 1784 (2010).

[18]T. Y. Choi, D. J. Hwang and C. P. Grigoropoulos, Appl. Surf. Sci. **197-198**, 720 (2002).

[19]I. H. Chowdhury, X. Xu and A. M. Weiner, Appl. Phys. Lett. **86**, 151110 (2005).

[20]S. Singha, Z. Hu and R. J. Gordon, J. Appl. Phys. **104**, 113520 (2008).

[21]Y. Ralchenko, A. E. Kramida and J. Reader, "NIST Atomic Spectra Database," NIST ASD Team, 2012. [Online]. Available: http://physics.nist.gov/asd. [Accessed 20 March 2013].

[22]A. Taflove and S. C. Hagness, *Computational Electrodynamics: The Finite-Difference Time-Domain Method*, 3rd ed. (Artech House, Norwood, MA, 2005).

[23]C. K. Birdsall and A. B. Langdon, *Plasma Physics via Computer Simulation* (Taylor and Francis Group, New York, 2005).

[24]R. W. Hockney and J. W. Eastwood, *Computer Simulation Using Particles* (McGraw-Hill, New York, 1981).

[25]R. Kupfer, B. Barmashenko and I. Bar, Phys. Rev. E **88**, 013307, (2013).

[26]K. S. Yee, IEEE Trans. Antennas Propag. **14**, 302 (1966).

[27]J. P. Boris, in *Proceedings of the 4th Conference on Numerical Simulation of Plasmas* (Naval Research Lab, Washington, D.C., 1970).

[28]T. Umeda, Y. Omura, T. Tominaga and H. Matsumoto, Comput. Phys. Commun. **156**, 73 (2003).

[29]X. Kong, M. C. Huang, C. Ren and V. K. Decyk, J. Comp. Phys. **23**0, 1676 (2011).

[30]P. Gibbon, *Short Pulse Interactions with Matter: An Introduction* (Imperial College, London, 2005).

[31]W. L. Kruer, *The Physics of Laser Plasma Interactions* (Addison-Wesley, New York, 1988).

[32]A. Ng, P. Celliers, A. Forsman, R. M. More, Y. T. Lee, F. Perrot, M. W. C. Dharma-wardana and G. A. Rinker, Phys. Rev. Lett. **72**, 3351 (1994).

[33]H. M. Milchberg, R. R. Freeman, S. C. Davey and R. M. More, Phys. Rev. Lett. **61**, 2364 (1988).

[34]D. von der Linde and H. Schuler, J. Opt. Soc. Am. B **13**, 216 (1996).

[35]K. G. Estabrook, E. J. Valeo and W. L. Kruer, Phys. Fluids **18**, 1151 (1975).

[36]R. A. Cairns, Plasma Phys. **20**, 991 (1978).

[37]F. David and R. Pellat, Phys. Fluids **23**, 1682 (1980).

[38]J. Hohlfeld, S.-S. Wellershoff, J. Gudde, U. Conrad, V. Jahnke and E. Matthia, Chem. Phys. **251**, 237 (2000).

[39]M. E. Povarnitsyn, T. E. Itina, K. V. Khishchenko and P. R. Levashov, Phys. Rev. Lett. **103**, 195002 (2009).